\documentclass[twocolumn,showpacs,aps,prl,superscriptaddress]{revtex4-1}
\usepackage{graphicx}
\usepackage{dcolumn}
\usepackage{amsmath}
\usepackage{epsfig}
\usepackage{subfloat}
\usepackage{colordvi}
\usepackage{hhline}
\usepackage{multirow}
\usepackage{subfigure}
\usepackage{rotating}
\usepackage{slashbox}
\usepackage{hyperref}
\usepackage{xcolor}
\usepackage{lineno}
\usepackage{url}

\RequirePackage{xspace}

\newcommand{\gev}{\ensuremath{\mathrm{\,Ge\kern -0.1em V}}\xspace}
\newcommand{\mev}{\ensuremath{\mathrm{\,Me\kern -0.1em V}}\xspace}
\newcommand{\gevc}{\ensuremath{\mathrm{\,Ge\kern -0.1em V\!/}c}\xspace}
\newcommand{\mevc}{\ensuremath{{\mathrm{\,Me\kern -0.1em V\!/}c}}\xspace}
\newcommand{\gevcc}{\ensuremath{{\mathrm{\,Ge\kern -0.1em V\!/}c^2}}\xspace}
\newcommand{\gevccs}{\ensuremath{{\mathrm{\,Ge\kern -0.1em V^2\!/}c^4}}\xspace}
\newcommand{\mevcc}{\ensuremath{{\mathrm{\,Me\kern -0.1em V\!/}c^2}}\xspace}

\def\epem       {\ensuremath{e^+e^-}\xspace}
\def\myX       {\ensuremath{X(1835)}\xspace}
\def\jpsi     {\ensuremath{J/\psi}\xspace}

\def\qqbar {\ensuremath{q\overline q}\xspace}

\def\c     {\ensuremath{c}\xspace}

\def\etap   {\ensuremath{\eta'}\xspace}
\def\pipi      {\ensuremath{\pi^+\pi^-}\xspace}

\def\ppbar     {\ensuremath{p\bar{p}}\xspace}

\def\etappipi     {\ensuremath{\eta'\pi^{+}\pi^{-}}\xspace}

\mathchardef\Upsilon="7107
\def\Y#1S{\ensuremath{\Upsilon{(#1S)}}\xspace}

\def\Bbar    {\kern 0.18em\overline{\kern -0.18em B}{}\xspace}

\def\kkmc     {\mbox{\tt KKMC}\xspace}
\def\lundcharm     {\mbox{\tt LUNDCHARM}\xspace}

\def\evtgen     {\mbox{\tt EvtGen}\xspace}
\def\geant      {\mbox{\tt GEANT4}\xspace}

\begin{document}

\title{
\Large \bf {\boldmath Measurement of the branching fraction of $\jpsi\rightarrow\omega\etap\pipi$ and search for
$\jpsi\rightarrow\omega\myX$, $\myX\rightarrow\etap\pipi$ decay}
} 
\author{
\begin{center}
M.~Ablikim$^{1}$, M.~N.~Achasov$^{10,d}$, P.~Adlarson$^{56}$, S. ~Ahmed$^{15}$, M.~Albrecht$^{4}$, M.~Alekseev$^{55A,55C}$, A.~Amoroso$^{55A,55C}$, F.~F.~An$^{1}$, Q.~An$^{52,42}$, Y.~Bai$^{41}$
% ...
\\
\vspace{0.2cm}
(BESIII Collaboration)\\
\vspace{0.2cm} {\it
$^{1}$ Institute of High Energy Physics, Beijing 100049, People's Republic of China\\
% ...
$^{4}$ Bochum Ruhr-University, D-44780 Bochum, Germany\\
% ...
$^{10}$ G.I. Budker Institute of Nuclear Physics SB RAS (BINP), Novosibirsk 630090, Russia\\
% ...
$^{15}$ Helmholtz Institute Mainz, Johann-Joachim-Becher-Weg 45, D-55099 Mainz, Germany\\
% ...
$^{41}$ Southeast University, Nanjing 211100, People's Republic of China\\
$^{42}$ State Key Laboratory of Particle Detection and Electronics, Beijing 100049, Hefei 230026, People's Republic of China\\
% ...
$^{52}$ University of Science and Technology of China, Hefei 230026, People's Republic of China\\
% ...
$^{55}$ (A)University of Turin, I-10125, Turin, Italy; (B)University of Eastern Piedmont, I-15121, Alessandria, Italy; (C)INFN, I-10125, Turin, Italy\\
$^{56}$ Uppsala University, Box 516, SE-75120 Uppsala, Sweden\\
% ...
$^{d}$ Also at the Novosibirsk State University, Novosibirsk, 630090, Russia\\
% ...
}\end{center}
\vspace{0.4cm}
}

\begin{abstract}
Using a sample of $1.31\times 10^{9}$ \jpsi events collected by the BESIII detector at  BEPCII during 2009 and 2012,
we study the $\jpsi\rightarrow\omega\etap\pipi$ hadronic process.
For the first time, we measure the branching ratio $\mathcal{B}(\jpsi\rightarrow\omega\etap\pipi)=(1.12\pm0.02\pm0.13)\times 10^{-3}$.
We search for the \myX state in the $\etap\pipi$ invariant mass spectra.
No evidence is found and we estimate the upper limit on the branching fraction at $90\%$ confidence level 
to be $\mathcal{B}(\jpsi\rightarrow\omega\myX, \, \myX\rightarrow\etap\pipi) < 6.2 \times 10^{-5}$.
\end{abstract}

\maketitle
\noindent

One of the main topics of the BESIII physics program is the search for unconventional hadronic states.
Among the light hadrons, the \myX state has caught the attention both from an experimental 
and a theoretical point of view. 
It was observed first in the $\eta' \pipi$ invariant mass spectra at BES in the $\jpsi\rightarrow\gamma\eta'\pipi$ 
radiative decay~\cite{PhysRevLett.95.262001},
and confirmed later with much higher statistics by BESIII~\cite{PhysRevLett.106.072002}. 
Its mass and width were measured to be $M=1836.5\pm3.0^{+5.6}_{-2.1}$ \mevcc and $\Gamma=190\pm9^{+38}_{-36}$\mev, 
with the product of branching fractions 
$\mathcal{B}(\jpsi\rightarrow\gamma\myX)\cdot\mathcal{B}(\myX\rightarrow\eta'\pipi)=(2.87\pm0.09^{+0.49}_{-0.52})\times10^{-4}$
 ~\cite{PhysRevLett.106.072002}.
 The \myX state was also seen in the process $\jpsi\rightarrow\gamma K_S^0K_S^0\eta$~\cite{PhysRevLett.115.091803};
 its mass and width were found to be in agreement with those measured in Ref.~\cite{PhysRevLett.106.072002}, and
 he quantum numbers $J^{PC}$ were determined to be $0^{-+}$ from a  partial wave analysis.

Just a few years before the observation of the \myX state, an anomalous enhancement close to the \ppbar mass threshold, called $X(1860)$,
 has been observed by BES in the $\jpsi\rightarrow\gamma\ppbar$ decay~\cite{PhysRevLett.91.022001},
 and confirmed by BESIII~\cite{Ablikim2010421} and CLEO~\cite{PhysRevD.82.092002},
 while no evidence has been seen in other channels, such as $\jpsi\rightarrow\omega\ppbar$~\cite{refId0,PhysRevD.87.112004}
 or $\jpsi\rightarrow\phi\ppbar$~\cite{PhysRevD.93.052010}.
 A partial wave analysis of the \ppbar mass-threshold enhancement was performed~\cite{PhysRevLett.108.112003}, 
 and the $J^{PC}$ quantum number  were determined to be the same as for the \myX.
The discovery of these new states has stimulated many theoretical speculations on their nature,
such as a \ppbar bound state~\cite{PhysRevC.80.045207,PhysRevD.80.034032,EPJA28.351}, 
a pseudo-scalar glueball~\cite{PhysRevD.74.034019,Kochelev2006283,Hao200653}, 
a radial excitation of the $\eta'$ meson~\cite{PhysRevD.73.014023}, {\it etc}.
Thanks to the world's largest $\epem\to\jpsi$ data set collected by BESIII, it has been possible to study in detail
the significant abrupt change in the line shape of the $\myX\rightarrow\eta'\pipi$ in correspondence 
of the $p\bar{p}$ mass threshold~\cite{PhysRevLett.117.042002},
which  could be originated from the opening of the \ppbar additional decay channel (threshold effect)
or by the interference between two different resonances. 
However, none of the hypotheses could be excluded and no final conclusion has been made.
In order to extract additional information about the states around $1.85$ \gevcc\
with the present BESIII statistics, additional decay modes must be investigated.

In this paper, we report on the search for \myX in the $\jpsi\rightarrow\omega\eta'\pipi$ process.
The comparison of the production rates between $\jpsi\rightarrow\omega\myX$ and $\jpsi\rightarrow\gamma\myX$
could also help to get information on the \qqbar or gluon component of \myX~\cite{PhysRevD.74.034019,EPJA28.351},
{\it i.e.}~if \myX contains substantial \qqbar components, like the $\eta'$ meson, it should be observed in $\jpsi\rightarrow\omega\myX$.
Using  the branching fraction of $\jpsi\rightarrow\omega(\phi)\eta'$, 
 the branching fraction of $\jpsi\rightarrow\omega(\phi)\myX$ is estimated to be in the order of $10^{-5}$~\cite{PhysRevD.74.034019}.
On the other hand, a very small branching fraction is expected for larger gluon component.  
Another estimation was done in Ref.~\cite{EPJA28.351}, where  $\mathcal{B}(\jpsi\rightarrow\omega\myX)$ 
is expected to be two orders of magnitude less than that of  $\jpsi\rightarrow\gamma\myX$ decay.

This analysis is based on $1.31\times10^9$  \jpsi events collected by BESIII
during 2009 and 2012.
The BESIII detector~\cite{Ablikim2010345} is a magnetic spectrometer operating at BEPCII,
a double-ring \epem collider with center-of-mass energies ranging from $2.0$ to $4.6$ \gev.
The geometrical acceptance covered is $93\%$ of a $4\pi$ solid angle.
From the inner to the outer side, it consists of a helium-based main drift chamber (MDC),
a time-of-flight system (TOF) and a CsI(Tl) electromagnetic calorimeter (EMC),
all enclosed in a superconducting solenoidal magnet providing a magnetic field of $1$ T ($0.9$ T in 2012).
The solenoid is surrounded by an octagonal flux-return yoke with resistive plate chambers
interleaved with steel.

A \geant-based Monte Carlo (MC) simulation package~\cite{Agostinelli2003250} is used to optimize selection criteria,
estimate background processes, and determine detection efficiency.
The production of the \jpsi resonance is simulated with \kkmc event generator~\cite{JADACH2000260,PhysRevD.63.113009},
while the decays are generated with \evtgen~\cite{LANGE2001152,1674-1137-32-8-001}.
Simulated inclusive \jpsi events of approximatively the equivalent luminosity of data are used 
to study background processes. 
The known decays of \jpsi are modeled with branching fractions being set to the world average values
from Particle Data Group (PDG)~\cite{Olive:2016xmw},
while the remaining decays are generated with \lundcharm~\cite{PhysRevD.62.034003}.
We simulate 700,000 MC events using phase space model for the processes
$\jpsi\rightarrow\omega\etap\pipi$ and $\jpsi\rightarrow\omega\myX, \, \myX\rightarrow\etap\pipi$,
which are used to optimize the event selection and to determine the selection efficiency.
For the $\jpsi\rightarrow\omega\myX$ signal simulation we also take into account the $J^{PC}=0^{-+}$
quantum numbers.

\begin{figure}[!htb]
\centering
 \includegraphics[width=0.48\textwidth] {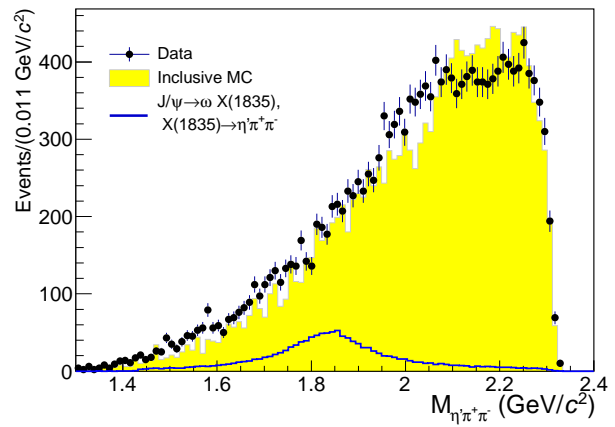}
 \caption{(color online). $\etappipi$ invariant mass distribution for data (black points), inclusive MC sample (yellow histogram)
 and $\jpsi\rightarrow\omega\myX, \, \myX\rightarrow\etappipi$ signal MC sample with an arbitrary normalization  (blue line).}
\label{fig:Xdistribution}
\end{figure}

For each candidate event, we select charged tracks well reconstructed in the MDC detector with the polar 
angle $\theta$ satisfying the condition $|\cos\theta|<0.93$.
The tracks are required to pass the interaction point within $\pm 10$ cm along the beam direction
and within $1$ cm in the plane perpendicular to the beams.
Photon candidates are reconstructed using clusters of energy deposited in the EMC.
The energy deposited in the TOF is also included in EMC measurements in order to improve 
the reconstruction efficiency and the energy resolution.
Good photon candidates are required to have a deposited energy larger than $25$ MeV in the barrel region ($|\cos\theta|<0.8$)
and $50$ MeV in the end caps ($0.86 <|\cos\theta|<0.92$).
To eliminate those clusters associated with charged tracks, the angle between the direction 
of any charged track and the photon candidate must be larger than $5^\circ$.
Clusters due to the electronic noise and energy deposit unrelated to the event are suppressed
by requiring the shower time to be within $700$ ns of the event start time.
Events with six charged tracks, net charge equal to zero, and at least four photon candidates 
that satisfy the above requirements are retained for further studies.

\begin{figure}[!htb]
\centering
 \includegraphics[width=0.48\textwidth] {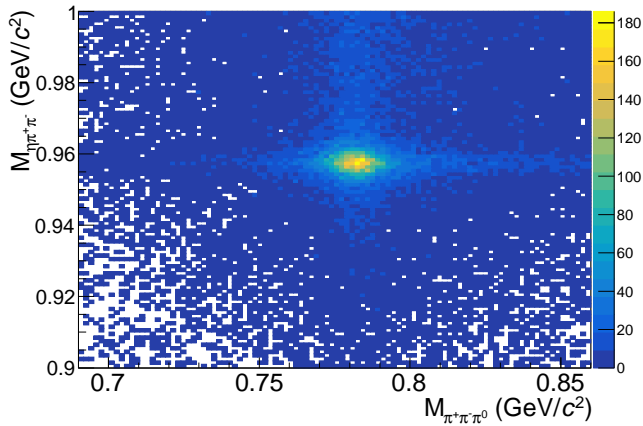}
 \caption{(color online). Scatter plot of $\eta\pipi$ invariant mass  as a function of $\pipi\pi^0$ invariant mass distribution.}
\label{fig:2dplot}
\end{figure}

In the reconstruction of $\jpsi\rightarrow\omega\etap\pipi$, the $\omega$ meson is reconstructed in its dominant $\pipi\pi^0$ decay mode 
and \etap via $\etap\rightarrow\eta\pipi$, while both $\eta$ and $\pi^0$ are reconstructed from $\gamma\gamma$ pairs
after applying the corresponding mass constrained kinematic fit.
To improve momentum resolution, for each $\pi^0\eta\pipi\pipi\pipi$ combination a four constraints ($4C$) energy-momentum kinematic fit is
performed. We select only events with $\chi^2_{4C}<60$.
In order to determine the $\pipi$ pairs produced in  $\omega/\etap$ decays, we select the combination which minimize the quantity
$\sqrt{(M_{\pipi\pi^0}-m_{\omega})^2+(M_{\eta\pipi}-m_{\etap})^2}$, where $m_{\omega}$ and $m_{\etap}$ are
the nominal masses of $\omega$ and $\etap$~\cite{Olive:2016xmw}, respectively,
while $M_{\pipi\pi^0}$ ($M_{\eta\pipi}$) is the $\pipi\pi^0$ ($\eta\pipi$) invariant mass.
Then, we require $M_{\pipi\pi^0}$ and $M_{\eta\pipi}$ to be within $3\sigma$ the fitted widths:
$|M_{\pipi\pi^0}-m_{\omega}|< 22$ \mevcc and  $|M_{\eta\pipi}-m_{\etap}|< 12$ \mevcc.
Figure~\ref{fig:Xdistribution} shows the $\etappipi$ invariant mass distribution ($M_{\etappipi}$) from data sample 
for those events that satisfy the selection criteria. No clear enhancement is visible.

In Fig.~\ref{fig:Xdistribution},  the $\etappipi$ invariant mass spectrum from inclusive MC sample
is also reported.
Since there are some discrepancies between the two distributions, more pronounced for $M_{\etappipi}>1.9$ \gevcc,
 we cannot use the inclusive MC sample
to model the background contribution. 
As an alternative, a two-dimensional fit to the $\pipi\pi^0$ and $\eta\pipi$ distributions will be used to 
get the number of $\jpsi\rightarrow\omega\etap\pipi$ signal events.
The scatter plot of $M_{\eta\pipi}$ as a function of $M_{\pipi\pi^0}$ is reported in Fig.~\ref{fig:2dplot}.
The $\omega$ signal is parametrized by a Breit-Wigner (BW) function convolved with a double Gaussian
and the $\etap$ signal by a double Gaussian function,  while third-order polynomial functions are used
for both $\omega$ and $\etap$ backgrounds.
All the parameters are treated as free with the exception 
of the $\omega$ width, which is fixed to the world average value~\cite{Olive:2016xmw}. 

\begin{figure*}[!ht]
\centering
\includegraphics[scale=0.7]{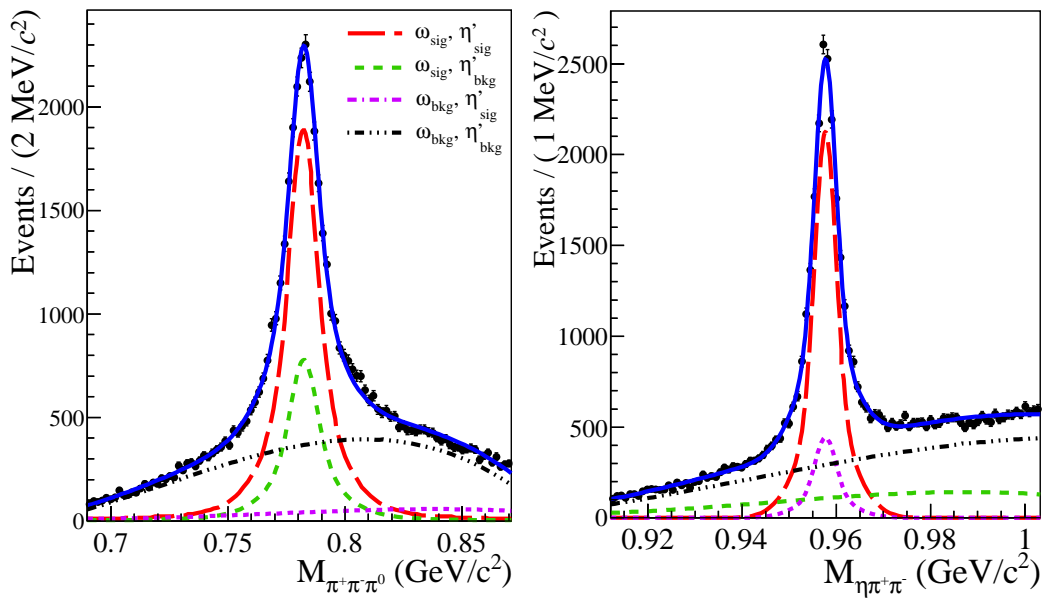}
\caption{ (color online). One-dimensional projections of two dimensional fit results to the $\pipi\pi^0$ (left) and $\eta\pipi$ (right) invariant mass distributions.
Blue curves refer to the final fit result, while the other fit components are represented by colored dashed curves:
red for $\omega$ and $\etap$ signals, green for $\omega$ signal and $\etap$ background,
magenta for $\omega$ background and $\etap$ signal, and black for both $\omega$ and $\etap$ backgrounds. 
}
\label{fig:2Dfit}
\end{figure*}

The one-dimensional projections of the fit result are shown in Fig.~\ref{fig:2Dfit}.
The branching fraction of the $\jpsi\rightarrow\omega\etap\pipi$ process is calculated with
\begin{equation}
\mathcal{B}(\jpsi\rightarrow\omega\etap\pipi)=\frac{N_{\rm{sig}}}{N_{\jpsi}\cdot\epsilon\cdot\mathcal{B}_{\rm{int}}},
\end{equation}
where $N_{\rm{sig}}=14151\pm287$ is the number of $\jpsi\rightarrow\omega\etappipi$ signal events from the fit to the data sample,
$N_{\jpsi}$ the number of $\jpsi$ events~\cite{Njpsi}, 
$\epsilon=6.48\%$  the detection efficiency calculated from signal simulation,
and $\mathcal{B}_{\rm{int}}$ the product of the decay branching fractions for the
$\omega\rightarrow\pipi\pi^0$, $\pi^0\rightarrow\gamma\gamma$, $\etap\rightarrow\eta\pipi$, and $\eta\rightarrow\gamma\gamma$
intermediate states quoted from PDG~\cite{Olive:2016xmw}.
The branching fraction is then determined to be 
$\mathcal{B}(\jpsi\rightarrow\omega\etap\pipi)=(1.12\pm0.02)\times10^{-3}$, 
where the uncertainty is statistical only.
 
\begin{figure}[!htb]
\centering
 \includegraphics[width=0.45\textwidth] {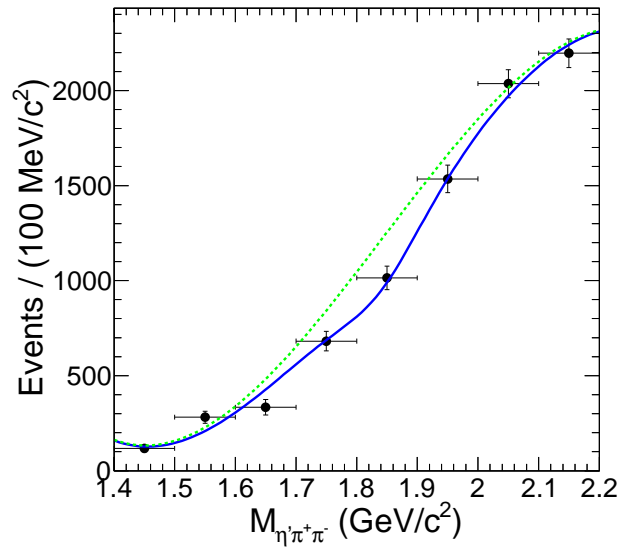}
 \caption{(color online). 
 $\chi^2$-fit result (blue curve) to the background subtracted $\etappipi$ invariant mass spectrum (black dots) 
 extracted as described in the text.
 Dashed green curve shows the background contribution which is parameterized 
 by means of a third-order polynomial function, while for the signal component we use 
 an efficiency-weighted BW function.}
\label{fig:flatte}
\end{figure}

As can been seen from Fig.~\ref{fig:Xdistribution}, no significant \myX signal is observed in the $\etappipi$ invariant mass spectrum,
and hence we extract the upper limit (UL) on the number of \myX signal events.
As stated before, we cannot use inclusive MC sample to parametrize the background shape.
Also a polynomial function may not be appropriate to describe a large background component under a very small
signal fraction of a broad resonance.
As an alternative in order to extract the background-corrected distribution, the two-dimensional fit 
to the $\pipi\pi^0$ and $\eta\pipi$ invariant mass spectra is performed
in eight slices of $\etappipi$ mass spectrum from $1.4$ \gevcc to $2.2$ \gevcc.
The background-subtracted $\etappipi$ invariant mass is shown in Fig.~\ref{fig:flatte}. 
The UL  on the number of $\myX$ signal events is extracted by means of a $\chi^2$-fit.
In this fit, all processes other than $\jpsi\rightarrow\omega\myX$ are considered as background,
and we assume there is no interference between \myX and non-\myX components.
Both the \myX signal and background yields are fitted as free parameters, and
we associate to the signal yield a Gaussian distribution with mean and width equal to
the number of signal events and the corresponding uncertainty resulting from the $\chi^2$-fit.
Then, the UL at $90\%$ confidence level (C.L.) is obtained 
by finding the point where the cumulative probability of this Gaussian distribution is equal to 0.9.

In the $\chi^2$-fit, two different signal functions are taken into account: 
a BW function with \myX mass and width fixed to values from Ref.~\cite{PhysRevLett.106.072002},
and a Flatt\'e function with fixed parameters from Ref.~\cite{PhysRevLett.117.042002},
both weighted by the efficiency,
while a third-order polynomial function is used for the background.
Systematic effects are evaluated by changing the $\etappipi$ fit range and the bin size,
as well as by varying the fit parameters within one standard deviation.
Since the $\etappipi$ background-corrected distribution is extracted 
from a two-dimensional fit to the $\pipi\pi^0$ and $\eta\pipi$ invariant mass spectra,
we need to evaluate its systematic contribution. 
 On this purpose,  three different signal functions are used to parametrize the $\omega$ and $\etap$ signal:
(1) a BW convolved with a double Gaussian for $\omega$ and a double Gaussian for $\etap$, 
(2) the $\omega$ and $\etap$ MC shapes, and 
(3) the convolution of the  $\omega$ and $\etap$ MC shapes with a double Gaussian.    
The resulting $\etappipi$ background-corrected distribution are  then fitted using 
a $\chi^2$-fit, as described before.
The fit that gives the largest result is then used to extract the UL on the number of \myX signal events 
at $90\%$ C.L., which amounts to $N^{UL}=582$.
The corresponding UL on the branching fraction of the $\jpsi\rightarrow\omega\myX, \, \myX\rightarrow\etappipi$ decay
at $90\%$ C.L. is calculated as
\begin{eqnarray}
\mathcal{B}&(\jpsi\rightarrow\omega\myX, \, \myX\rightarrow\etappipi) < \nonumber \\
&\frac{N^{UL}}{N_{\jpsi}\cdot\epsilon'\cdot\mathcal{B}_{\rm{int}}\cdot(1-\sigma_{\rm{sys}})} = 6.2 \times 10^{-5},
\end{eqnarray}
where $\epsilon'=5.26 \%$ is the \myX selection efficiency in the $\omega-\etap$ signal region,
and $\sigma_{\rm{sys}}$ is the total systematic uncertainty reported in Table~\ref{tab:syst}
and discussed below.

Several sources of systematic uncertainties are considered:
uncertainty due to the total number of $\jpsi$ events~\cite{Njpsi},
intermediate branching fractions~\cite{Olive:2016xmw}, data-MC differences in tracking efficiency,
photon detection efficiency, selection efficiencies, angular distributions, kinematic fit, signal and background functions and fit range.
Uncertainties due to the tracking efficiency for charged tracks are 
determined using control samples of $\jpsi\rightarrow\pipi p\bar{p}$ and $\jpsi\rightarrow K_S^0K^\pm\pi^\mp$.
The difference between the tracking efficiency in data and MC simulations is $1\%$ for each charged track. 
However, since we have six charged pions in the final state, and hence pions with very low momentum,
we check for possible tracking efficiency underestimation. 
We correct our MC simulations according to the data, also taking into account possible difference in the polar angle distributions,
and we find a tracking efficiency consistent with $6\%$.
For the neutral candidates, control samples of $\jpsi\rightarrow\rho\pi^0$ and $e^+e^-\rightarrow\gamma\gamma$
are used to study the photon detection efficiency, which amount to $1\%$ for each photon candidate.

\begin{table}[htp]
\small
\begin{center}
\caption{\normalsize{Summary of systematic uncertainties. 
Those items marked with ``-" has been taken into account in obtaining the 
 UL on the number of \myX signal events.}}
\begin{tabular}{c|c|c}
\hline
\hline
Sources & $\mathcal{B}(\jpsi\rightarrow\omega\etap\pipi)$ & UL \\
 & ($\%$) & ($\%$) \\
 \hline
  Number of $\jpsi$ &0.5 &0.5 \\
 \hline
  $\mathcal{B}_{\rm{int}}(\omega\rightarrow\pi^0\pipi)$ &0.78 &0.78 \\
  $\mathcal{B}_{\rm{int}}(\pi^0\rightarrow\gamma\gamma)$ & 0.03&0.03 \\
  $\mathcal{B}_{\rm{int}}(\etap\rightarrow\eta\pipi)$ &1.63 &1.63 \\
  $\mathcal{B}_{\rm{int}}(\eta\rightarrow\gamma\gamma)$ &0.51 &0.51 \\
 \hline
  Tracking efficiency &6 &6 \\
 \hline
 Photon detection &4 &4 \\
 \hline
 Selection efficiency &3.2 & negligible \\
 \hline
Angular distribution & 1.0 & - \\
\hline
 Kinematic fit & 5 & 5\\
 \hline
 $\omega/\etap$ signal function &4.8&-\\
 \hline
Fit range  &2.6 & - \\
\hline
Background shape &4.3 & - \\
\hline \hline
Total & 11.8 &9.0 \\
\hline
\hline
\end{tabular}
\label{tab:syst}
\end{center}
\end{table} 

The systematic contributions related to the selection efficiency used to calculate both branching 
fraction and upper limit are evaluated by means of additional MC samples, 
in which also different intermediate states are considered.
However, since no obvious structures are observed in the different combinations of
two- or three-particles invariant mass distributions,
we simulate a MC sample, without intermediate resonances, taking into account the spin-parity of the initial
and final states, and the difference in the efficiency is taken as systematic contribution.
Additional contribution can arise from differences between the angular distribution
of data and simulation in the $\jpsi\rightarrow\omega\etap\pipi$ process. We simulate a new MC sample following the same angular dependence as in the data.
The difference in the efficiency amount to $1\%$, which is taken as systematic uncertainty.

A control sample of $\psi(3686)\rightarrow\pipi\jpsi, \, \jpsi\rightarrow \pipi\pi^0\etap, \, \etap\rightarrow \eta\pipi$
is used to determine the systematic uncertainty related to the kinematic fit.
We perform a 2-dimensional fit to the $\jpsi$ and $\etap$ invariant mass spectra in order to extract
the number of signal events and calculate the efficiency as a function of $\chi^2_{4C}$.
The difference between data and MC in correspondence of the $\chi^2_{4C}$ cut used in this analysis
is taken as systematic uncertainty.  

Systematic contributions associated only with the branching fraction are those
related to the two-dimensional fit of the $\pipi\pi^0$ and $\eta\pipi$ invariant masses.
In particular, the systematic related to the signal functions are evaluated by means of MC shape 
distributions for both the $\omega$ and $\etap$ invariant mass spectra.
For the background, instead, we change the order of the polynomial function.
In both cases, the difference in the number of signal events is taken as systematic uncertainty.
We also change the fit range by a step of $5$ \mevcc, and the difference in the signal yield is taken as
 systematic uncertainty. 
 
Table~\ref{tab:syst} summarizes all sources of systematic uncertainties, for which the total contribution is obtained as 
sum of them in quadrature.

Using a sample of $1.31\times10^{9}$ \jpsi events collected with the BESIII detector,
we measure for the first time  the branching fraction for the decay  $\jpsi\rightarrow\omega\etappipi$
to be $(1.12\pm0.02\pm0.13)\times 10^{-3}$,
where the first uncertainty is statistical and the second systematic.
We also search for the \myX state in the hadronic \jpsi decay $\jpsi\rightarrow\omega\myX$,
with $\myX\rightarrow\etappipi$. 
No significant signal is observed and the upper limit at $90\%$ C.L. on the 
branching fraction is determined to be 
$\mathcal{B}(\jpsi\rightarrow\omega\myX, \, \myX\rightarrow\etappipi) < 6.2\times 10^{-5}$. 
Since the \myX state is observed only in radiative \jpsi decays and the branching fraction
is measured to be of the order of $10^{-4}$~\cite{PhysRevLett.106.072002,PhysRevLett.115.091803, PhysRevLett.117.042002},
authors of Ref.~\cite{PhysRevD.74.034019} suggest that a smaller branching fraction measured in hadronic \jpsi decays
could be an indication of a large gluon component.
Authors of Ref.~\cite{EPJA28.351} treat \myX as a baryonium with sizable gluon content, and estimate a
branching ratio of the order of $10^{-6}$.
Unfortunately, our upper limit result is too large to confirm or distinguish among several theoretical interpretations,
but it provides the first search for the \myX state in  \jpsi hadronic decays, 
which can be further investigated by studying additional hadronic decay modes.

\section*{Acknowledgment}
The BESIII collaboration thanks the staff of BEPCII and the IHEP computing center for their strong support. This work is supported in part by National Key Basic Research Program of\
 China under Contract No. 2015CB856700; National Natural Science Foundation of China (NSFC) under Contracts Nos. 11335008, 11425524, 11625523, 11635010, 11735014; the Chinese Acade\
my of Sciences (CAS) Large-Scale Scientific Facility Program; the CAS Center for Excellence in Particle Physics (CCEPP); Joint Large-Scale Scientific Facility Funds of the NSFC and\
 CAS under Contracts Nos. U1532257, U1532258, U1732263; CAS Key Research Program of Frontier Sciences under Contracts Nos. QYZDJ-SSW-SLH003, QYZDJ-SSW-SLH040; 100 Talents Program o\
f CAS; INPAC and Shanghai Key Laboratory for Particle Physics and Cosmology; German Research Foundation DFG under Contracts Nos. Collaborative Research Center CRC 1044, FOR 2359; I\
stituto Nazionale di Fisica Nucleare, Italy; Koninklijke Nederlandse Akademie van Wetenschappen (KNAW) under Contract No. 530-4CDP03; Ministry of Development of Turkey under Contra\
ct No. DPT2006K-120470; National Science and Technology fund; The Swedish Research Council; U. S. Department of Energy under Contracts Nos. DE-FG02-05ER41374, DE-SC-0010118, DE-SC-\
0010504, DE-SC-0012069; University of Groningen (RuG) and the Helmholtzzentrum fuer Schwerionenforschung GmbH (GSI), Darmstadt.

\bibliography{mybiblio}{}

\end{document}